\documentclass[conference]{IEEEtran}
\IEEEoverridecommandlockouts
\usepackage{cite}
\usepackage{amsmath,amssymb,amsfonts}
\usepackage{algorithmic}
\usepackage{graphicx}
\usepackage{subcaption}
\usepackage{booktabs}
\usepackage{adjustbox}
\usepackage[table,xcdraw]{xcolor}
\usepackage{textcomp}
\usepackage{multirow}
\usepackage{multicol}
\usepackage{booktabs}
\usepackage{bigstrut}
\usepackage{verbatim}
\usepackage{array}
\usepackage{rotating}
\usepackage{xcolor}
\usepackage{dblfloatfix}
\usepackage{soul}
\usepackage{gensymb}
\usepackage[a4paper, total={184mm,239mm}]{geometry}

\begin{document}

\title{qPRO-AQFP: Post-Routing Optimization of AQFP Circuits with Delay Line Clocking}
\author{Robert S. Aviles*, Ziyu Liu*, Jingkai Hong,\\ 
Sasan Razmkhah, Massoud Pedram~\IEEEmembership{IEEE Fellow}, Peter A. Beerel~\IEEEmembership{Senior Member, IEEE}

\thanks{*R. S. Aviles and Z. Liu contributed equally to this work and are considered co–first authors.}\thanks{This work has been supported by ARL DEVCOM under the FSDL: ColdPhase project, grant number W911NF-24-1-0317.

The authors are with the Department of Electrical and Computer Engineering,
University of Southern California, Los Angeles, CA 90007 USA (e-mail:
rsaviles@usc.edu; zliu4130@usc.edu, jingkaih@usc.edu, razmkhah@usc.edu, pedram@usc.edu, pabeerel@usc.edu)}}


\maketitle

\begin{abstract}
Adiabatic Quantum-Flux-Parametron (AQFP) logic is an ultra-low-power superconducting logic family with energy consumption approaching the Shannon limit, making it attractive for quantum computing control and cryogenic computing systems. Traditional AQFP designs face significant physical design challenges due to strict gate-level clocking requirements and limited interconnect lengths, leading to substantial buffer overhead and difficult timing closure. Recently, delay-line clocking of AQFP has been proposed to improve timing margins and reduce latency by enabling more flexible clock scheduling. However, prior work has primarily focused on placement and latency minimization, while relying on fixed timing parameters that do not capture the frequency dependence of AQFP setup and hold constraints.
To address this limitation, we propose a frequency-aware post-routing optimization framework that jointly optimizes clock period, latency, and timing slack under user-specified weighting. Experimental results across common benchmarks achieve 100\% post-routing timing closure across a range of performance–latency–slack trade-offs. Our approach also automates phase-skipping, reducing path-balancing buffer insertion by 34\% on average while only reducing operating frequency by 4\%.  
\end{abstract}


\section{Introduction}
Interest in emerging device technologies for application-specific low-power computation has increased due to CMOS power bottlenecks. Among these technologies, superconductor electronics (SCE) is particularly promising due to its ultra-low power operation, high switching speed, and compatibility with cryogenic environments~\cite{SFQ_EDA}. These properties make superconductor logic attractive for in-cryostat quantum computer control~\cite{DigiQ}, as well as in-fridge computation such as accelerated error correction~\cite{SFQ_ECC}.
Beyond quantum computing, SCE has also been demonstrated in classical workloads including binary neural networks~\cite{BNN_AQFP2023}, microprocessors~\cite{MANA}, and cryptographic accelerators~\cite{SCE-NTT}.  

Among SCE technologies, Adiabatic Quantum Flux Parametron (AQFP) logic~\cite{AQFP} is particularly attractive due to its ultra-low power consumption. AQFP circuits exhibit zero static power consumption and, even accounting for cryogenic cooling overhead, can achieve up to two orders of magnitude improvement in energy-delay product (EDP) compared to CMOS~\cite{Chen2019}. However, despite these advantages, superconducting technology imposes unique architectural and timing constraints that hinder large-scale integration and require specialized electronic design automation (EDA) solutions~\cite{fourie2023results}.

In particular, AQFP imposes two fundamental design constraints. \textbf{Gate-level clocking:} Every AQFP cell, including fanout (splitter) cells, must be driven by an AC clock. \textbf{Limited drive strength:} Logic gates can only drive short interconnect distances.
These constraints are traditionally addressed through clocked buffer insertion. Data transfer between cells requires overlapping clock phases, and reconvergent paths with mismatched logic depths require path-balancing buffers. In addition, buffer rows are inserted during placement to pipeline long interconnects. As a result, AQFP circuits often suffer from high buffer overhead. 

Conventional AQFP uses four clock phases with globally fixed offsets. More recently, delay-line clocking was proposed~\cite{takeuchi2019low} that uses a single meandering clock with optimizable delays between rows. This enables more flexible clock scheduling and reduces latency. However, prior work focused on placement and minimizing latency, does not explore design trade-offs, and does not capture the frequency-dependence of AQFP cell timing parameters~\cite{DLPlace}. 

Separately, phase-skipping has been proposed to reduce buffer overhead by enabling data transfer across non-consecutive clock phases~\cite{Nphaseclk}. Although demonstrated in fabricated designs such as an 8-bit ripple-carry adder~\cite{DL_RCA8}, the technique lacks any automated design support. 

To address these challenges, we present \textbf{qPRO-AQFP}, a post-routing optimization framework for AQFP designs that performs timing optimization and buffer reduction. Our key contributions are:
\begin{itemize}
    \item A post-routing clock delay scheduling optimization algorithm that allows trade-offs among target frequency, latency, and timing slack, while explicitly modeling the frequency dependence of cell parameters governing setup and hold constraints.
    \item A globally optimal AQFP buffer removal algorithm that achieves an average 34\% reduction in buffer count, improving manufacturability, and reducing latency by 5\%\, with only a 4\% throughput penalty on average.
    \item The first fully automated framework for timing-closed, post-routed AQFP circuits supporting both phase-skipping and delay-line clocking. 
\end{itemize}
The remainder of this paper is organized as follows. Section II reviews AQFP logic, clocking strategies, and physical design challenges. Section III presents our buffer-removal optimization and clock-scheduling algorithm. Section IV discusses experimental results, including comparisons 
to prior works, and Section V concludes the paper.

\section{Background}

\subsection{AQFP Logic and Timing}

All AQFP cells are derived from the fundamental buffer cell, which is driven by an AC clock line that adiabatically transitions the device 
into the logic state "0’ or ‘1". The resulting state produces an output current whose direction encodes the logic value. This basic buffer cell can be coupled to form inverters, splitter cells for fanout, constant logic sources, and three-input majority gates. 

Because every AQFP cell is clocked, each gate is subject to setup and hold timing constraints. 
Unlike conventional CMOS designs, however, AQFP does not use a single-phase global clock; instead, data can propagate between connected gates only when the relative clock arrival times satisfy the setup and hold constraints.


More specifically, Fig.~\ref{fig:abTiming} illustrates data transmission between two cells and highlights the relevant timing parameters. Notice that the clock at the gate $b$ may arrive significantly later than the clock at the gate $a$. Similarly to other logic families that employ delayed-clock propagation~\cite{isvlsi3,DelayBalancing}, the setup time therefore sets the \textit{lower bound} on the allowable difference between the clock arrival times at the gates and the \textit{clock delay}, while the hold time sets the \textit{upper bound}.

In AQFP, the clock-to-output delay ($c2q$), the setup time, and the hold time are influenced by the slope of the AC clock waveform, while the reset delay ($rd$) depends on the clock period, thus creating frequency-dependent timing parameters~\cite{AQFP_Timing}. 

More formally, for two connected gates $(i,j)$ with the gate $i$ driving the gate $j$, the setup constraint is expressed as
\begin{equation}\label{eq:Gensetup}
    clk_i + c2q_i + prop_{ij} \leq clk_j - setup_j
\end{equation}
\noindent where $clk_i$ and $clk_j$ denote the clock arrival times at gates i and j. This constraint enforces the minimum phase difference ($clk_j-clk_i$) and affects the minimum circuit latency, but not the clock period.

This hold constraint is expressed as
\begin{equation}\label{eq:Genhold}
    clk_i + c2q_i + prop_{ij} + rd_i  \geq clk_j + hold_j
\end{equation}
\noindent where $rd_i$ denotes the duration of the signal pulse on the gate output of $i$. 
If the sink clock ($clk_j$) arrives too late relative to the source clock ($clk_i$), the source output may reset before the sink samples it, causing a hold violation. When the relative clock skew ($clk_j-clk_i$) cannot be further reduced, this violation can be resolved by extending $rd_i$, which is achieved by increasing the clock period. In this way, the hold constraint effectively determines the maximum achievable clock frequency in AQFP circuits.

While AQFP designs are traditionally constrained to be fully path balanced through buffer insertion, sufficiently small clock skews can allow data transmission across imbalanced paths (a phenomenon known as \textit{phase-skipping})~\cite{Nphaseclk}. Fig~\ref{fig:DLSkip} illustrates one of such paths between the pair of gates $(c,d)$. Phase-skipping can reduce buffer overhead by 47\%~\cite{AvilesNPhase}, but introduces additional timing complexity that must be addressed during clock scheduling.

\subsection{Delay Line Clocking}

\begin{figure}[htbp]
\begin{subfigure}[t]{0.245\textwidth}
\centering
\includegraphics[width=\linewidth]{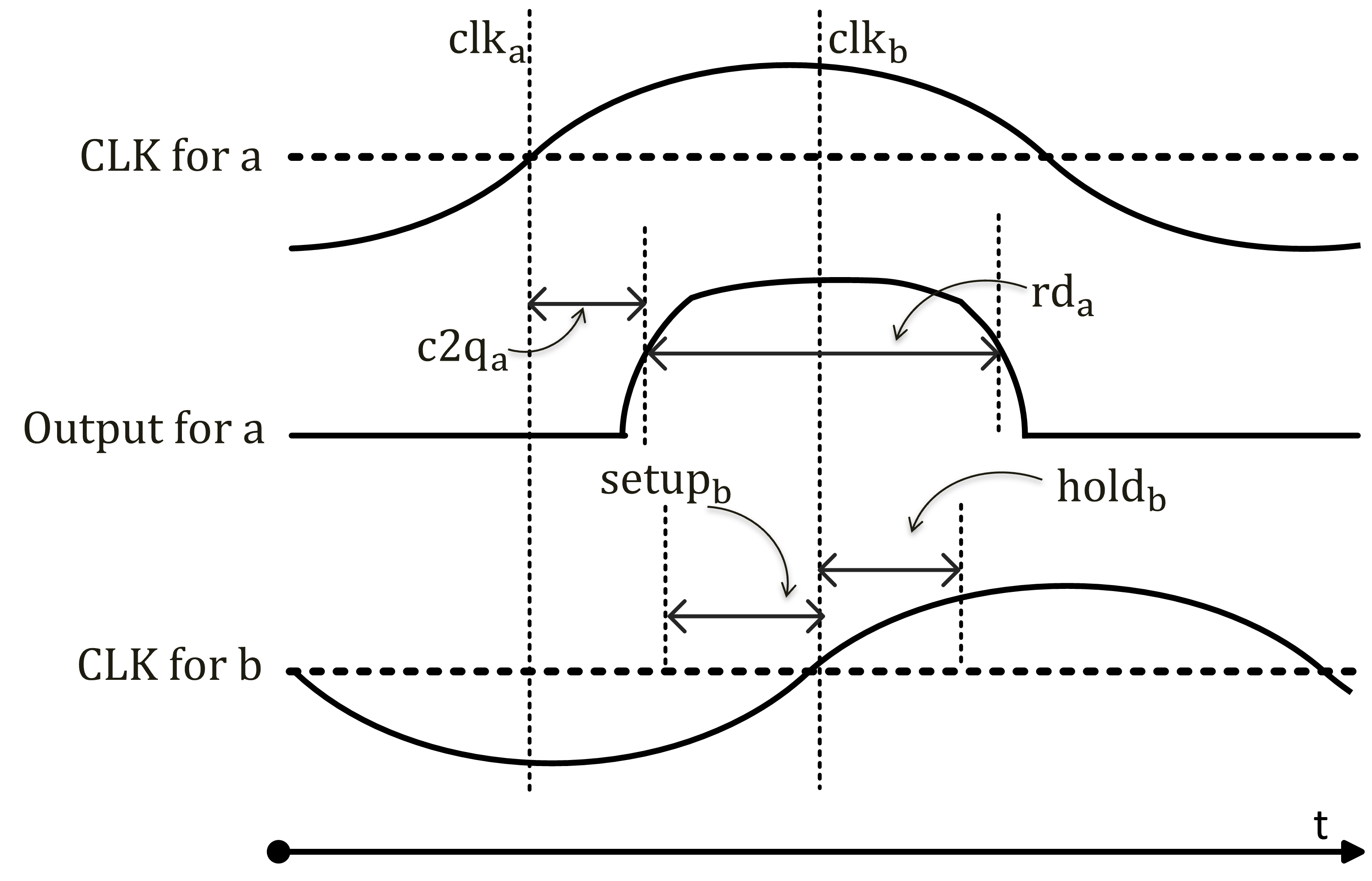}
\caption{$(a,b)$ timing parameters}\label{fig:abTiming}
\end{subfigure}%
\begin{subfigure}[t]{0.245\textwidth}
\centering
\includegraphics[width=\linewidth]{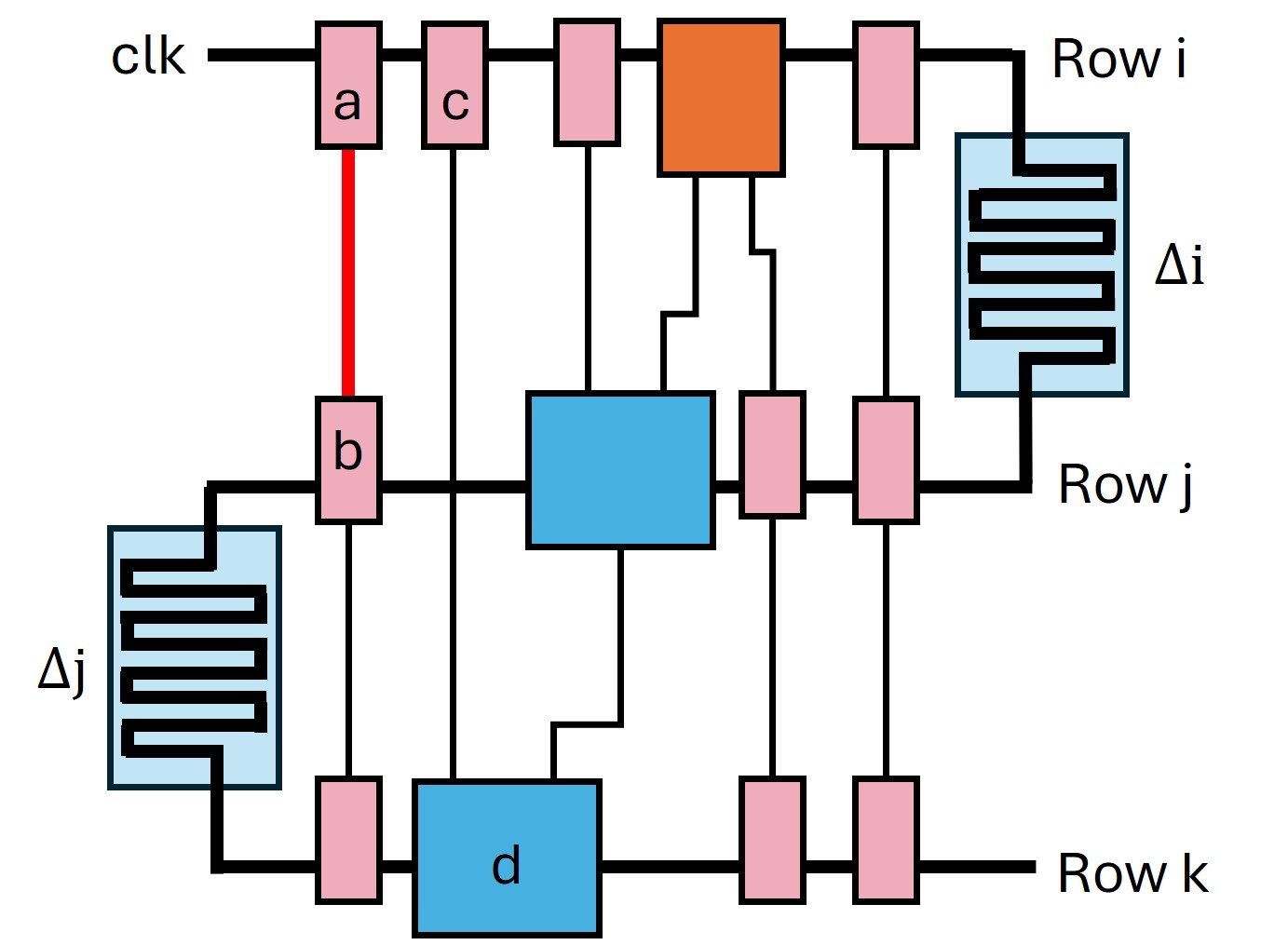}
\caption{Delay-line clock routing}\label{fig:DLSkip}
\end{subfigure}%
\caption{Delay line clocking and timing. $\Delta_{i}$ and $\Delta_{j}$ must be set to not only satisfy setup and hold of all row-to-row connections like $(a,b)$ but also any phase-skipping connections like $(c,d)$.}
\end{figure}
Delay-line clocking creates phase offsets between logic rows by routing the clock signal through serpentine interconnect segments, as illustrated in Fig.~\ref{fig:DLSkip}. The added wire length introduces a deliberate clock skew between rows, determined by clock-scheduling optimization. This approach contrasts with conventional four-phase AQFP clocking, where phase offsets are globally fixed. In delay-line clocking, the phase difference between rows can instead be tuned by adjusting clock routing lengths, allowing greater flexibility in satisfying setup and hold constraints and enabling techniques such as phase-skipping.

Previous work (DLPlace)~\cite{DLPlace} introduced a placement algorithm for delay-line clocked AQFP circuits in which clock delays between rows were determined during placement using estimated wire lengths and fixed timing parameters. However, these estimates may become inaccurate after routing, where interconnect delays can deviate significantly from half-perimeter wirelength (HPWL) estimates.

Additionally, it is important to clarify the hold constraint used in DLPlace~\cite{DLPlace}, reproduced below using the notation of Eq.~(\ref{eq:Genhold}):

\begin{equation}\label{eq:DLHold}
    (clk_j - clk_i) - prop_{ij} \leq T - hold_j + c2q_i
\end{equation}
This formulation replaces the reset delay $rd_i$ with the clock period $T$. Because $rd_i$ is typically only a fraction of the positive clock interval (e.g., $rd_i = 72$ ps for $T = 200$ ps) ~\cite{TAAS,AQFP_Timing}, DLPlace significantly relaxes the hold constraint and may lead to overly optimistic performance estimates. 

\section{Our Approach}

\textbf{qPRO-AQFP} is a post-routing optimization framework that first maximizes the removal of buffers while respecting interconnect constraints using a shortest path formulation and subsequently optimizes the clock delay schedule based on post-routing interconnect delays. In particular, this step uses an MILP formulation with user-defined optimization weights for clock period, latency, and timing slack, in which binary variables capture piecewise-linear models for frequency-dependent cell parameters. 





\begin{figure}[htbp] 
\begin{subfigure}[t]{0.245\textwidth}
\centering
\includegraphics[width=\linewidth]{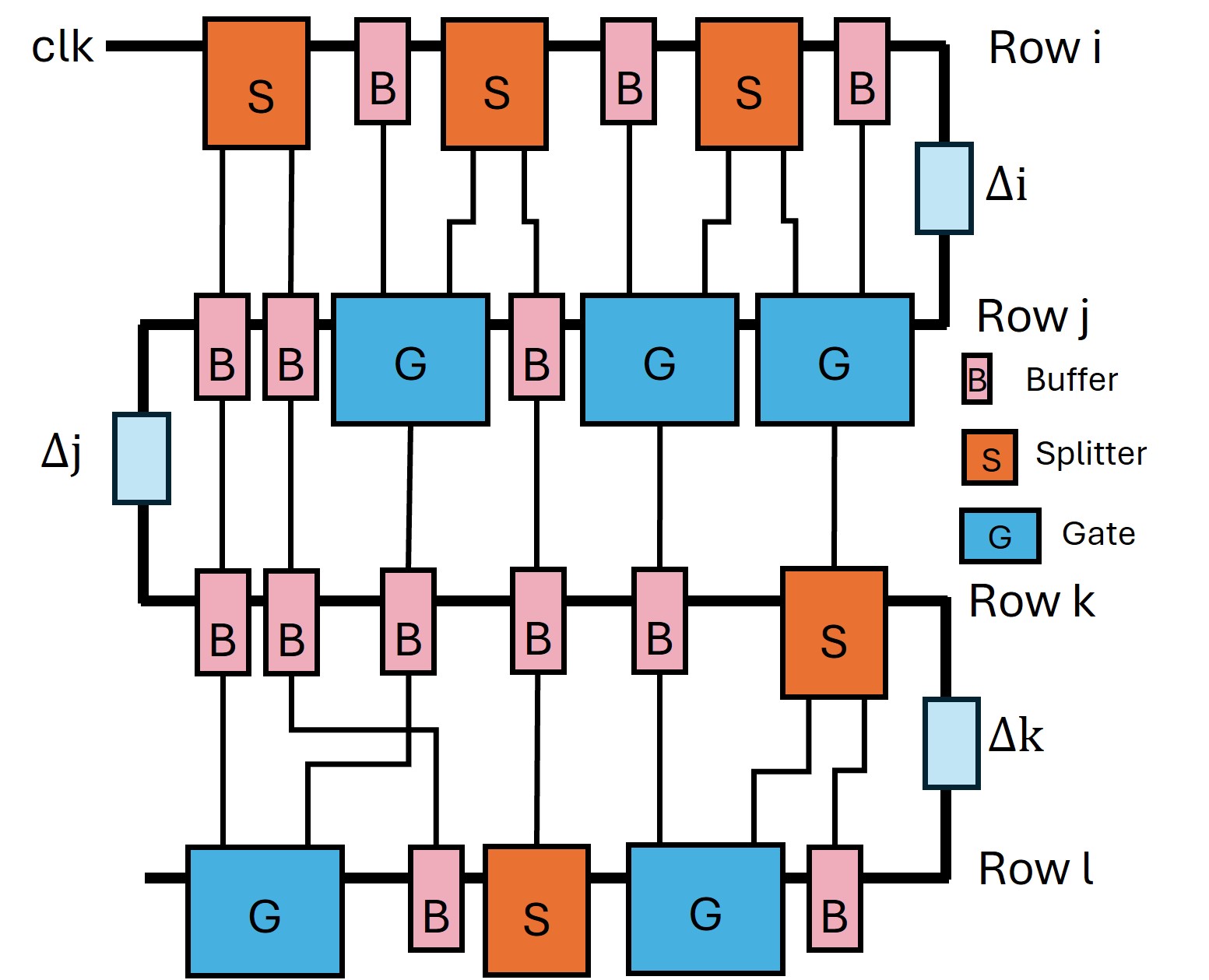}
\caption{Path-balanced circuit.}\label{fig:NoCompression}
\end{subfigure}%
\begin{subfigure}[t]{0.245\textwidth}
\centering
\includegraphics[width=\linewidth]{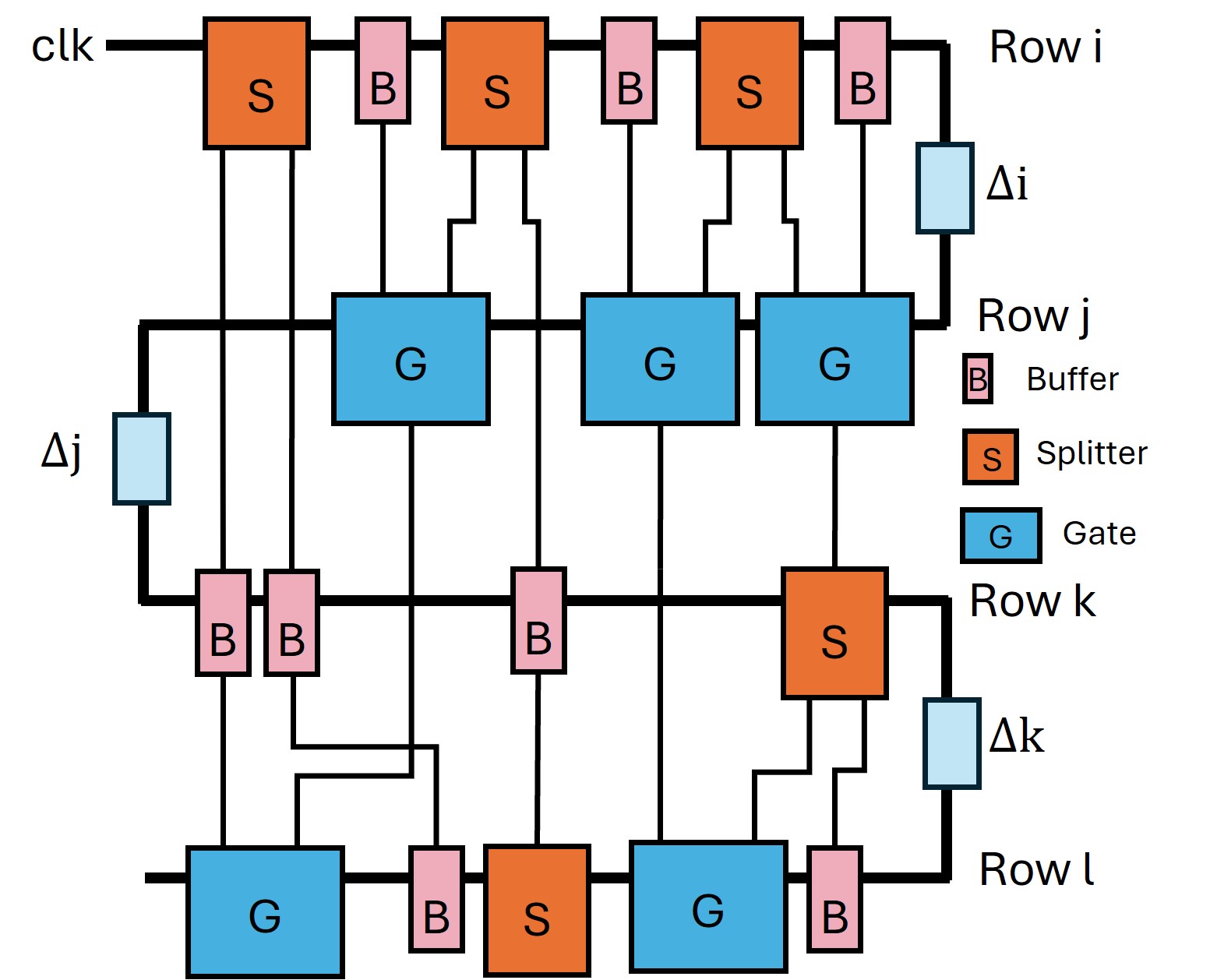}
\caption{Circuit w/ buffers removed.}\label{fig:Compressed}
\end{subfigure}
\caption{Illustrative impact of phase-skipping AQFP designs.}
\vspace{-2em}
\end{figure}

\subsection{Buffer Removal}

After the initial placement and routing stages, interconnect delay information is parsed to direct buffer removal. In general, a buffer can be removed only if the following interconnect length constraint is satisfied:
\begin{equation}
L_{\text{source} \rightarrow \text{buffer}} 
+ L_{\text{buffer}} 
+ L_{\text{buffer} \rightarrow \text{sink}}
\leq L_{\text{max\_drive}}
\label{eq:length_constraint}
\end{equation}
\noindent where $L_{\text{source} \rightarrow \text{buffer}}$ and $L_{\text{buffer} \rightarrow \text{sink}}$ denote the routed wire lengths extracted from the post-route layout database. $L_{\text{max\_drive}}$ is the maximum propagation distance at which the signal can still be reliably sampled without causing sampling errors~\cite{AQFP_interconnect}. $L_{\text{buffer}}$ represents the intrinsic pin-to-pin distance between the input and output pins of the buffer cell (which is non-trivial due to the layout size of the AQFP cells in current fabrication processes).


When a buffer is removed, the connection extends directly from the source to the sink along the original routed wire segments, with the internal pin-to-pin distance of the eliminated buffer incorporated into the route. 
Note that removing an intermediate buffer within a buffer chain may affect the feasibility of removing adjacent buffers. To systematically address this dependency, the buffer removal problem for each buffer chain is formulated as a maximum-weight source-to-sink path problem on a directed acyclic graph(DAG) as follows.
Consider a buffer chain comprising $m$ buffers. A directed acyclic graph $G = (V, E)$ is constructed with $V = \{v_0, v_1, \dots, v_m, v_{m+1}\}$, where $v_0$ and $v_{m+1}$ represent the source and sink of the buffer chain, respectively, and $v_1, \dots, v_m$ correspond to the physical buffer locations along the chain. A directed edge $(v_i, v_j) \in E$ with $0 \le i < j \le m+1$ is introduced if all buffers between $v_i$ and $v_j$ can be removed while satisfying the driving constraint:
\begin{equation}
L_{i \rightarrow j} \leq L_{\text{max\_drive}}.
\end{equation}
\noindent where $L_{i \rightarrow j}$ denotes the total length of pin-to-pin connection after removing the intermediate buffers, computed according to
the maximum propagation distance constraint defined in Eq.~\eqref{eq:length_constraint}. 
For each edge $(v_i, v_j)$, we define its weight as $w_{i,j} = j-i-1$, which corresponds to the number of intermediate buffers eliminated by selecting this edge. The optimal buffer removal solution is then obtained by identifying a path from the source node $v_0$ to the sink node $v_{m+1}$ that maximizes the cumulative edge weight:
\begin{equation}
\max_{P \in \mathcal{P}_{0\to m+1}} \sum_{(v_i,v_j) \in P} w_{i,j}
\end{equation}
\noindent where $\mathcal{P}_{0\to m+1}$ denotes the set of all feasible source-to-sink paths in $G$. Because $G$ is acyclic, this can be solved in linear time $O(|V|+|E|)$ using dynamic programming based on topological ordering as detailed below.


In AQFP, all fanouts are sourced by a splitter and terminated at a gate, ensuring that no inter-chain coupling constraints exist. Thus, the global objective decomposes as the sum of the per-chain objectives. Formally, let $\mathcal{C}$ denote the set of buffer chains and let $P_c$ denote a source-to-sink path for chain $c \in \mathcal{C}$. Since the global objective is additive,
\begin{equation}
\max_{P_{\text{all}}} \sum_{c\in\mathcal{C}} \sum_{(v_i,v_j)\in P_c} w_{i,j}
\end{equation}
\noindent and the feasible set factorizes into independent per-chain feasible sets, then the global problem decomposes as
\begin{equation}
\max_{P_{\text{all}}} \sum_{c \in \mathcal{C}} \sum_{(v_i,v_j) \in P_c} w_{i,j}
= \sum_{c \in \mathcal{C}} \max_{P_c} \sum_{(v_i,v_j) \in P_c} w_{i,j}
\end{equation}
Thus, the longest-path problem is solved independently for each chain, yielding the globally optimal buffer removal configuration when inter-chain interactions are absent.  

\subsection{Post-Routing Clock Schedule Optimization}

After routing and any buffer removal, precise delay values are fixed for all nets. These delay values are then used to construct the timing constraints for all connections in the design. If $i$ and $j$ are placed in adjacent rows $m$ and $n$, respectively, then the clock difference can be simplified to:
\begin{equation}
    clk_j- clk_i = \Delta clk_{ij} + \Delta m
\end{equation}
\noindent where $\Delta clk_{ij}$ is the base delay of the clock traveling from $i$ through the row $m$ and through the row $n$ to $j$, and $\Delta m$ is the additional delay introduced after the row $m$ by the serpentine clock routing between rows, as shown in Fig.\ref{fig:DLSkip}.  

Our setup and hold constraints in Eqs.~(\ref{eq:Gensetup})-(\ref{eq:Genhold}) can then be reformulated with all variables on the left-hand side as:
\begin{align}
    \Delta m - (c2q_i + setup_j) \geq prop_{ij} - \Delta clk_{ij} \\
    \Delta m - (c2q_i + rd_i - hold_j) \leq prop_{ij} - \Delta clk_{ij}  
\end{align}

For phase-skipping connections, we need modified constraints to account for additional row delays. For phase-skipping path $(i,k)$ where $i$ is in row $m$ and $k$ is in row $o$ with row $n$ being skipped, the constraints become:
\begin{align}
    \Delta m + \Delta n - (c2q_i + setup_k)\geq prop_{ik} - \Delta clk_{ik} \\
    \Delta m + \Delta n- (c2q_i + rd_i - hold_k) \leq prop_{ik} - \Delta clk_{ik} 
\end{align}

Given that cell timing parameters vary as a function of the clock period $T$ and that AQFP designs have a limited number of cell types (e.g., {buffer, majority3, splitter2, splitter3, splitter4}), we can define linearized regions for each cell type $i$ in set $C$. For example, $rd$ can be defined as $\forall i \in C$
\begin{equation}
    rd_i(T) = \begin{cases}
    a_iT + b_i & 0 \leq T \leq A \\
    c_iT + d_i & A < T \leq B \\
    . \\
    . \\
    
    x_iT + y_i & Y < T \leq Z \\
    \end{cases}
\end{equation}
\noindent where the number of linearized regions can be increased to more closely approximate the exact behavior as needed.

Although this approach can support unique parametrized definitions for every cell type, we note that in typical cell libraries, all cells are derived from the buffer cell, which makes the timing parameter differences between cells typically very small \cite{AQFP_Timing}. 
As such, for simplicity and without loss of robustness, we may treat the timing parameters of each cell uniformly, while enforcing an appropriate minimum slack.

With this simplification, instead of enumerating each cell's parameters, we can reduce our setup and hold equations to combine all frequency-dependent timing parameters affecting setup and hold into $F_s(T)$ and $F_h(T)$, respectively.
\begin{align}
    F_S(T) = c2q(T) + setup(T) \\
    F_H(T) = c2q(T) + rd(T) - hold(T)
\end{align}

With these constraints formalized, we present the following MILP formulation, which takes user-specified parameters $\alpha, \beta,$ and $\gamma$ to prioritize the clock period (T), latency (L), and timing slack (S) in the objective function:

\begin{equation}
\text{Minimize:} \quad \tau T - \sigma S + \lambda L
\end{equation}
\begin{equation*}
\text{subject to:}
\end{equation*}
For all $(i,j)\in E$ such that $j\in n$, $i\in m$, and $n=m+1$:
\begin{equation}
\Delta_{m} - F_H + S 
\le 
prop_{ij} - \Delta clk_{ij} 
\label{eq:hold1}
\end{equation}
\begin{equation}
\Delta_{m} - F_S - S 
\ge 
Prop_{ij} - \Delta clk_{ij} 
\label{eq:setup1}
\end{equation}
For consecutive rows $(m,n,o)$ with $n=m+1$ and $o=n+1$:
\begin{equation}
\Delta_{m} + \Delta_{n} - F_H + S 
\le 
Prop_{ij} - \Delta clk_{ij} 
\label{eq:hold2}
\end{equation}
\begin{equation}
\Delta_{m} + \Delta_{n} - F_S - S 
\ge 
Prop_{ij} - \Delta clk_{ij} 
\label{eq:setup2}
\end{equation}
\begin{equation}
L - \sum_{m} \Delta_{m} = 0
\label{eq:length}
\end{equation}
\begin{equation}
S_{\min} \le S \le S_{\max}
\label{eq:Sbounds}
\end{equation}
\begin{equation}
T_{\min} \le T \le T_{\max}
\label{eq:Tbounds}
\end{equation}
\begin{equation}
F_H = a T_1 + b z_1 
     + c T_2 + d z_2 
     + e T_3 + f z_3
\label{eq:FH}
\end{equation}
\begin{equation}
F_S = g T_1 + h z_1 
     + i T_2 + j z_2 
     + k T_3 + \ell z_3
\label{eq:FS}
\end{equation}
\begin{equation}
T = T_1 + T_2 + T_3
\label{eq:Tsum}
\end{equation}
\begin{equation}
z_1 + z_2 + z_3 = 1
\label{eq:zsum}
\end{equation}
\begin{equation}
0 \le T_1 \le A z_1
\label{eq:T1}
\end{equation}
\begin{equation}
A z_2 < T_2 \le B z_2
\label{eq:T2}
\end{equation}
\begin{equation}
B z_3 < T_3 \le T_{\max}\cdot z_3
\label{eq:T3}
\end{equation}
\begin{equation}
\Delta_{m},\, T_i,\, S,\, L,\, F_H,\, F_S \in \mathbb{R}
\end{equation}
\begin{equation}\label{eq:bin}
z_1, z_2, z_3 \in \{0,1\}
\end{equation}

\begin{table*}[tb]
\caption{qPRO-AQFP results demonstrating post-routing timing-closed circuits without violations across 
the design space. In particular, $\tau$, $\lambda$, and $\sigma$ denote the timing, latency, and slack objective weighting, respectively.}
\label{tab:priority}

\centering
\setlength{\tabcolsep}{3.5pt}
\renewcommand{\arraystretch}{1.15}

\begin{tabular}{l ccc ccc ccc}
\toprule

& \multicolumn{3}{c}{$\tau \gg \lambda \gg \sigma$}
& \multicolumn{3}{c}{$\tau \gg \lambda \gg \sigma$ ($S_{\min}=5$)}
& \multicolumn{3}{c}{$\tau \gg \sigma \gg \lambda$} \\

\cmidrule(lr){2-4}
\cmidrule(lr){5-7}
\cmidrule(lr){8-10}

\textbf{Circuit} 
& \textbf{Freq. (GHz)} & \textbf{Latency (ps)} & \textbf{Slack (ps)}
& \textbf{Freq. (GHz)} & \textbf{Latency (ps)} & \textbf{Slack (ps)}
& \textbf{Freq. (GHz)} & \textbf{Latency (ps)} & \textbf{Slack (ps)} \\

\midrule

\texttt{adder8}
& 5.0 & 570  & 0
& 5.0 & 734  & 5
& 5.0 & 1223 & 19.8 \\

\texttt{apc32}
& 5.0 & 596  & 0
& 5.0 & 807  & 5
& 5.0 & 1546 & 22.6 \\

\texttt{apc128}
& 4.9 & 3275 & 0
& 4.3 & 3598 & 5
& 3.5* & 4372 & 14.6 \\

\texttt{c432}
& 5.0 & 1376 & 0
& 5.0 & 1778 & 5
& 5.0 & 2946 & 19.2 \\

\texttt{c499}
& 4.4 & 2861 & 0
& 3.9 & 3073 & 5
& 3.5* & 3369 & 10.3 \\

\texttt{c1355}
& 4.5 & 2876 & 0
& 4.0 & 3063 & 5
& 3.5* & 3397 & 11.1 \\

\texttt{c1908}
& 5.0 & 2725 & 0
& 5.0 & 3270 & 5
& 5.0 & 3961 & 11.3 \\

\texttt{sorter32}
& 4.7 & 2422 & 0
& 4.2 & 2684 & 5
& 3.5* & 3215 & 13.5 \\

\bottomrule
\end{tabular}

\vspace{3pt}
\raggedright

\footnotesize{$^{*}$Frequency in these designs limited to 3.5 GHz to enable slack and latency trade-offs to be visible.}
\vspace{-1em}
\end{table*}

Here, $S_{min}$ is the lower bound on the timing slack defined by the user, and $S_{max}$ is the upper bound on the benefit of slack to our objective function defined by the user.
Eqs.~(\ref{eq:FH})-(\ref{eq:T3}) model the piecewise linearization of cell timing characteristics. Constants A, B, and $T_{max}$ define the bounds of the linearized region of timing characterization. The binary variables $z_1, z_2, z_3$ ensure that only 1 linearized segment is activated at a time, while Eqs.~(\ref{eq:T1})-(\ref{eq:T3}) ensure that the correct segment is selected and the others are forced to 0. For example, consider when $T = A$, then $z_1$ will be set to $1$, and all other binary variable will be 0, forcing $T_2 = 0$ and $T_3 = 0$. Then $T = T_1 + 0$ and $F_H = aT_1 + bz_1 + 0$.

Note that AQFP timing parameters are fairly linearizable with respect to clock period \cite{AQFP_Timing}; as such, the number of binary variables required to model the various segments is relatively small. The number of binary variables is independent of the number of cells in the library and the circuit size, allowing our approach to scale to large circuits. Furthermore, for clarity, we assumed uniform $F_H$ and $F_S$ for all cells; however, for cell libraries with widely varying parameter constraints, (\ref{eq:FH})-(\ref{eq:FS}) is replaced with a similar linearization with unique constants for each cell parameter (without additional binary variables required). 

\section{Experimental Results}

This section presents experimental results for the proposed qPRO-AQFP optimization framework. All circuits were generated using qPALACE synthesis and physical design tools using cell-library timing parameters derived from JoSim simulations. 



Table~\ref{tab:priority} demonstrates a range of cells we used to test qPRO-AQFP under different optimization priorities. 
Because clock period, latency, and timing slack often conflict, we used weights along with minimum and maximum bounds defined in Eqs.~(\ref{eq:Sbounds})-(\ref{eq:Tbounds}). In the first configuration ($\tau \gg \lambda \gg \sigma$), the optimization prioritizes minimizing the clock period (subject to the maximum supported library frequency of 5 GHz), and then minimizes circuit latency at that period. In the second configuration, a slack constraint of $S_{min}=5$ restricts the solution space to allow 5 ps of slack across all timing constraints and then, in order of priority, minimizes the clock period and latency. 
The final configuration demonstrates a design scenario in which the slack is maximized after the maximum operating frequency is reached. 

Table~\ref{tab:sta_param_impact} further illustrates the space exploration design capabilities of qPRO-AQFP using the \texttt{adder8} benchmark. For example, the configuration $\tau \gg \sigma \gg \lambda$, with $S_{\max}=10$ps first ensures the lowest clock period is achieved and then optimizes the slack without regard for the impact of latency up to 10 ps first minimizes the clock period, then increases the slack up to 10 ps without regard for latency, after which latency is minimized. 

The MILP-based timing optimization is implemented using the Python package \texttt{scipy.optimize.milp}. All experiments were executed on an EPYC 7763 Milan CPU. For all scheduling optimizations, runtimes were negligible, with solution times below 0.1 seconds.

Importantly, all our designs successfully closed timing, demonstrating that qPRO-AQFP efficiently enables timing closure while supporting flexible trade-offs among performance, latency, and timing slack for delay-line-clocked AQFP circuits.

\begin{table}[tb]
\caption{Impact of qPRO-AQFP optimization parameters on
\texttt{adder8}.}
\label{tab:sta_param_impact}

\centering
\setlength{\tabcolsep}{2pt}
\renewcommand{\arraystretch}{1.15}
\begin{adjustbox}{width=\columnwidth}
\begin{tabular}{l c c c}
\toprule
\textbf{Parameter Setting} & \textbf{Frequency (GHz)} & \textbf{Latency (ps)} & \textbf{Min Slack (ps)} \\
\midrule

$\tau \gg \lambda \gg \sigma$ 
& 5.0 & 570  & 0.0 \\

$\tau \gg \lambda \gg \sigma$, ($S_{min}=5$)
& 5.0 & 734  & 5.0 \\

$\tau \gg \lambda \gg \sigma$, ($S_{min}=25$)
& 4.4 & 1395 & 25.0 \\
$\tau \gg \sigma \gg \lambda$ 
& 5.0 & 1223 & 19.8 \\

$\tau \gg \sigma \gg \lambda$, ($S_{\max}=10$)
& 5.0 & 900  & 10.0 \\

$\tau \gg \sigma \gg \lambda$, ($T_{min}=250$)
& 4.0 & 1520 & 28.8 \\

\bottomrule
\end{tabular}
\end{adjustbox}
\end{table}

\subsection{Phase-Skipping Impacts}

In this final section, we present the impact of phase-skipping buffer removal within qPRO-AQFP, as shown in Table~\ref{tab:qpro_ps}. For four of the circuits, peak performance is maintained, while the remaining four require a reduction in operating frequency to satisfy hold constraints caused by increased clock path length. On average, this corresponds to only a 4\% decrease in frequency in exchange for a 5\% reduction in latency and a 34\% reduction in buffer overhead. The reduction in buffer overhead lowers the junction count in our benchmarks by 22\%, leading to a comparable reduction in power consumption and simplifying fabrication, potentially improving yield since fewer junctions are present.

\begin{table}[tb]
\caption{Impact of Phase-Skipping within qPRO-AQFP.}
\label{tab:qpro_ps}

\centering
\setlength{\tabcolsep}{4pt}
\renewcommand{\arraystretch}{1.15}
\begin{adjustbox}{width=\columnwidth}

\begin{tabular}{l cc cc c}
\toprule
\multirow{2}{*}{\textbf{Circuit}} 
& \multicolumn{2}{c}{\textbf{Baseline}} 
& \multicolumn{2}{c}{\textbf{Phase-Skipping}} 
& \multirow{2}{*}{\textbf{Buf. Saved (\%)}} \\
\cmidrule(lr){2-3}
\cmidrule(lr){4-5}

& Freq. (GHz) & Lat. (ps)
& Freq. (GHz) & Lat. (ps)
& \\
\midrule

\texttt{adder8}     & 5.0 & 570  & 5.0 & 539  & 31.66 \\
\texttt{apc32}    & 5.0 & 596  & 5.0 & 552  & 38.69 \\
\texttt{apc128}   & 4.9 & 3275 & 4.5 & 3192 & 35.04 \\
\texttt{c432}     & 5.0 & 1376 & 5.0 & 1247 & 42.14 \\
\texttt{c499}     & 4.4 & 2861 & 4.0 & 2785 & 36.93 \\
\texttt{c1355}    & 4.5 & 2876 & 4.1 & 2836 & 37.36 \\
\texttt{c1908}    & 5.0 & 2725 & 5.0 & 2458 & 37.15 \\
\texttt{sorter32} & 4.7 & 2422 & 4.3 & 2394 & 16.50 \\
\midrule
\rowcolor[rgb]{.9,.9,.9}
\multicolumn{3}{l}{\textbf{Average Percentage Change}} 
& \textbf{-4.3\%} 
& \textbf{-5.0\%} 
& \textbf{-34.4\%} \\
\bottomrule
\end{tabular}
\end{adjustbox}
\end{table}

\begin{figure}[t]
\includegraphics[angle=90,width=\linewidth]{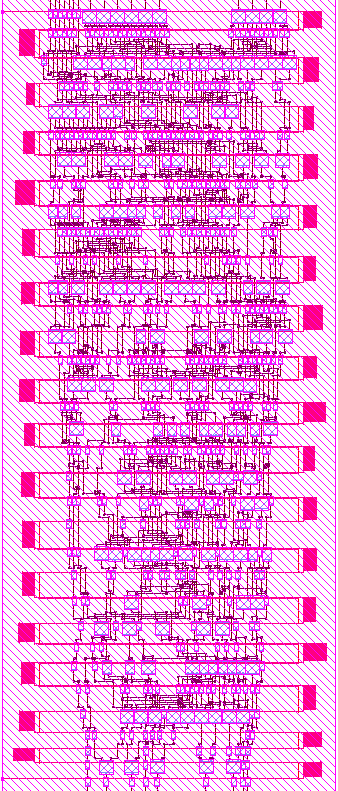}
\centering
\caption{8-bit adder utilizing phase-skipping for buffer reduction. Layout rotated {90\degree} for convenience with primary inputs on the left.}\label{fig:adder8}
\vspace{-2em}
\end{figure}

We show in Fig.~\ref{fig:adder8} a fully automated, post-routing, timing closed 8-bit adder utilizing phase-skipping for buffer reduction.

\begin{table*}[hptb]
\caption{Post-Routing Results Compared to State-of-the-Art Placement Results.}
\label{tab:placement_comparison}

\centering
\setlength{\tabcolsep}{4pt}
\renewcommand{\arraystretch}{1.15}

\begin{tabular}{l ccc ccc ccc}
\toprule
\multirow{2}{*}{\textbf{Circuit}} 
& \multicolumn{3}{c}{\textbf{TAAS \cite{TAAS}}} 
& \multicolumn{3}{c}{\textbf{DLPlace \cite{DLPlace}}} 
& \multicolumn{3}{c}{\textbf{qPRO-AQFP}} \\

\cmidrule(lr){2-4}
\cmidrule(lr){5-7}
\cmidrule(lr){8-10}

& Freq (GHz) & Latency (ps) & $rd$ (ps) 
& Freq (GHz)$^{*}$ & Latency (ps) & $rd$ (ps)
& Freq (GHz) & Latency (ps) & $rd$ (ps) \\

\midrule

\texttt{adder8}      & 6.1 & 2009 & $\leq$ 82 & 5 & 594  & 200 & 5.0$^{**}$ & 570  & 72 \\
\texttt{apc32}       & 5.8 & 2279 & $\leq$ 86 & 5 & 625  & 200 & 5.0$^{**}$ & 596  & 72 \\
\texttt{apc128}      & 3.9 & 9936 & $\leq$ 128 & 5 & 2401 & 200 & 4.9 & 3275 & 74 \\
\texttt{c432}        & 5.4 & 3608 & $\leq$ 93 & 5 & 1310 & 200 & 5.0$^{**}$ & 1376 & 72 \\
\texttt{c499}        & 4.3 & 5407 & $\leq$ 116 & 5 & 2926 & 200 & 4.4 & 2861 & 82 \\
\texttt{c1355}       & 4.1 & 5427 & $\leq$ 122 & 5 & 2631 & 200 & 4.5 & 2876 & 81 \\
\texttt{c1908}       & 4.4 & 5966 & $\leq$ 114 & 5 & 2681 & 200 & 5.0 & 2725 & 72 \\
\texttt{sorter32}    & 4.7 & 3138 & $\leq$ 106 & 5 & 2046 & 200 & 4.7 & 2422 & 76 \\

\bottomrule
\end{tabular}

\vspace{3pt}
\raggedright
\footnotesize{$^{*}$DLPlace sets $rd$ = T in the hold constraint, which significantly relaxes timing requirements.
$^{**}$Frequency constrained to 5 GHz by the adiabatic nature of the cell library.}
\vspace{-2em}
\end{table*}
\subsection{Comparison to State of the Art Placement Results}

For completeness, Table~\ref{tab:placement_comparison} compares our results with prior art.
However, note that our results are
post-routing compared to placement-only results from prior work, a significant benefit of our approach. Moreover, we note that these prior works use different cell libraries with distinct timing parameters and drive strengths, further complicating direct comparison.  In particular, the results reported in TAAS include operating frequencies above 5 GHz, which exceed the adiabatic operating range of our cell library. 
Under this constraint, qPRO-AQFP achieves lower latency and comparable or better performance than TAAS on several benchmarks.
Comparison with DLPlace is further complicated by differences in the hold constraint formulation. DLPlace replaces the reset delay term with the clock period ($rd = T$), as shown in Equation~\ref{eq:DLHold}. In practice, reset delay is only a fraction of the clock period, and increasing $T$ is typically required to increase $rd$ and avoid hold violations. To illustrate this discrepancy, we add our $rd$ to the table, highlighting that our work achieves equal or near performance to DLPlace while having much tighter timing constraints. Since TAAS does not report $rd$ directly, but lists its constraint as $rd=\frac{1}{4}T -2\cdot c2q$, we list an approximation of $\frac{1}{4}T$ for their results. For qPRO-AQFP, we report our actual cell-library reset delay values corresponding to the selected clock frequency. 
Despite these methodological differences and inconsistencies, the comparisons suggest that our post-routing timing-closed circuits achieve performance comparable to state-of-the-art placement-stage results.

\section{Conclusions}
In this work, we present qPRO-AQFP, the first fully automated framework for generating timing-closed, post-routed AQFP circuits supporting both delay-line clocking and phase-skipping.
%
%
Our delay-line clocking optimization explicitly captures the frequency dependence of cell parameters that govern setup and hold constraints, enabling correct timing optimization across a range of operating points with runtimes of less than 0.1 seconds. Our buffer removal algorithm is globally optimal and, on average, achieves a 34\% reduction in buffer count, a 22\% reduction in junction count, and a 5\% decrease in latency, with only a modest drop in operating frequency.
Although direct comparisons with prior placement-based approaches are limited by differences in timing assumptions, closed-source tools, and proprietary cell libraries, our post-routing results achieve performance comparable to state-of-the-art placement-level methods while providing more realistic timing evaluation.
Together, these contributions establish qPRO-AQFP as an effective framework for timing closure and architectural optimization of AQFP circuits, helping bridge the gap between placement-level optimization and physically realizable superconducting logic designs. Future work can explore phase-skipping optimization during placement, as well as supporting advanced AQFP architectural schemes that enable \textit{phase-alignment} \cite{phaseMatch,avilesPS_PA}.
\bibliographystyle{IEEEtran}
\bibliography{bibliography}

\end{document}